\documentclass[3p,times,twocolumn]{elsarticle}

\usepackage{amssymb}

\usepackage[figuresright]{rotating}

\def \om {\Omega_{0m}}

\def \beq  {\begin{equation}}
\def \eeq  {\end{equation}}
\def \ber  {\begin{eqnarray}}
\def \eer  {\end{eqnarray}}

\begin{document}

\begin{frontmatter}




\title{Falsifying Cosmological Constant}

\author{Arman Shafieloo}

\address{Asia Pacific Center for Theoretical Physics, Pohang, Gyeongbuk 790-784, Korea \\
Department of Physics, POSTECH, Pohang, Gyeongbuk 790-784, Korea \\
email: arman@apctp.org}

\begin{abstract}
One of the main goals of physical cosmology is to reconstruct the expansion
history of the universe and finding the actual model of dark energy. In this article I review the difficulties of understanding dark energy and discuss about two strategic approaches, 'reconstructing dark energy' and 'falsifying dark energy models'. While one can use the data to reconstruct the expansion history of the universe and the properties of dark energy using novel approaches, considering the data limitations and its uncertainties we have to deal with cosmographic degeneracies that makes it difficult to distinguish between different dark energy models. On the other hand one can use the power of the data to falsify an assumed model using advanced statistical techniques. Within all these issues, focusing on falsification of cosmological constant has a particular importance since finding any significant deviation from $\Lambda$ would result      to a break through in theoretical physics, ruling out the standard concordance model of cosmology.

\end{abstract}


\end{frontmatter}


\section{Introduction}

The standard model of cosmology is based on the general theory of relativity. Einstein’s discovery of general relativity enabled us to develop a theory of the universe which is testable and can be falsified. So cosmology has become a proper science which can predict events and explain observations. The Big Bang model of the universe which is based on general relativity and is in fact the standard model of the universe at present, has successfully passed several important tests include the expansion of the universe as exhibited by the Hubble diagram; light element abundances which are in concordance with Big Bang nucleosynthesis predictions; observations of the cosmic microwave background which is a black body radiation left over from the young universe when the latter was only a few hundred thousand years old, etc. The standard cosmological model also needs to account for the origins of inhomogeneities such as galaxies, stars and planets. In the early 1980’s the inflationary model was suggested~\cite{inflation1,inflation2,inflation3} and subsequently shown to be able to successfully seed galaxy formation~\cite{structure1,structure2,structure3,structure4}. Besides the key issues of seed initial conditions for galaxies, the standard model must also account for dark matter and dark energy. Currently it is felt that the dark components of the universe, dark matter and dark energy, constitute around $96\%$ of the total energy density of the universe. However, it could also be that the presence of an unseen component implies a crises for the standard model of cosmology and calls for a revision of the general theory of relativity as advocated by some researchers. To determine which is the correct direction for theory to take one must develop sophisticated statistical methods and apply these to observational data in order to get a bias free picture of cosmological observations. The standard model of cosmology, known as `Vanilla model' because of its simplicity can be summarised as a spatially flat homogeneous and isotropic on large scales Friedman-Lometre-Robertson-Walker (FLRW) universe with power-law form of the primordial spectrum for the initial fluctuations and constitute of baryonic matter, cold dark matter and cosmological constant as dark energy. The standard model of cosmology is in fact based of many assumptions that leaves us with only 6 parameters to explain the universe and its dynamic. $\Omega_{bm}$ and $\Omega_{dm}$ (baryonic and dark matter densities) are two of these parameters and $\Omega_{\Lambda}=1-(\Omega_{bm}+\Omega_{dm})$ since the universe is assumed to be spatially flat. $\tau$ represents the epoch of reionization, $H_0$ the Hubble constant at $z=0$, $n_s$ the spectral index of the primordial fluctuations and $A_s$ the overall amplitude of the initial fluctuations are the other 4 key parameters. Out of all these, the first 4 parameters govern the dynamic of the universe and the background evolution and the last two represent the initial condition through power spectrum given by $P_R(k)=A_s[\frac{k}{k_*}]^{n_s-1}$ ($k_*$ is just a pivot point). One should admit that despite of simplicity of the concordance model, most cosmological observations are in good agreement with this model and in fact there is no strong evidence against it at current status of observations. However, agreement of this model with most cosmological observations does not necessarily mean that we have found the actual model of the universe. Different assumptions of the standard model can be independently tested using different statistical methods applied on various cosmological data, e.g, look at ~\cite{cmss,lp1} for testing the isotropy of the universe,~\cite{ag} for testing the structure formation suggested by the standard model, ~\cite{chris1,bruce1,smoothing3} for testing flatness of the universe and ~\cite{hss} for testing the power-law form of the primordial spectrum. In this article we focus on one the important aspects of the standard model of cosmology, namely, cosmological constant as dark energy. We first discuss about reconstructing dark energy and its difficulties using parametric or non-parametric approaches. Then we overview the theoretical cosmographic degeneracies between different cosmological quantities. Next we discuss about a different strategic approach to assume and falsify $\Lambda$ using cosmological observations rather than reconstructing the universe and at the end we conclude.
\section{Reconstructing Dark Energy}
The cosmological constant was originally introduced by Einstein in 1917~\cite{einstein}. The cosmological constant has a constant equation of state of $w = -1$. Although introduced in 1917 the so called $\Lambda$-term has had a checkered history. Its recent prominence in cosmological literature is largely due to supernovae observations~\cite{supernovae1,supernovae2} (look at~\cite{supernovae3} for the most recent compilation) supported by CMB and other datasets~\cite{lss_obs1,lss_obs2,lss_obs3}, which points to the interesting fact that the universe may currently be accelerating. To generate this acceleration one requires a component with a negative pressure and with a relatively large value of the energy density as compared with dark and baryonic matter. Currently it is our belief that the cosmological constant density must be as least $2/3$ of the total energy budget of the universe. However many other theoretical candidates for a matter component with similar characteristics to the cosmological constant have been proposed~\cite{de_candidates1,de_candidates2,de_candidates3,de_candidates4}. All these candidates together are called dark energy. One of the challenges of cosmology is to define which one is in fact responsible for the acceleration of the universe.

\begin{figure*}[!t]
\vspace{-0.8in}
\hspace{0.in}
\includegraphics[scale=0.50, angle=-90]{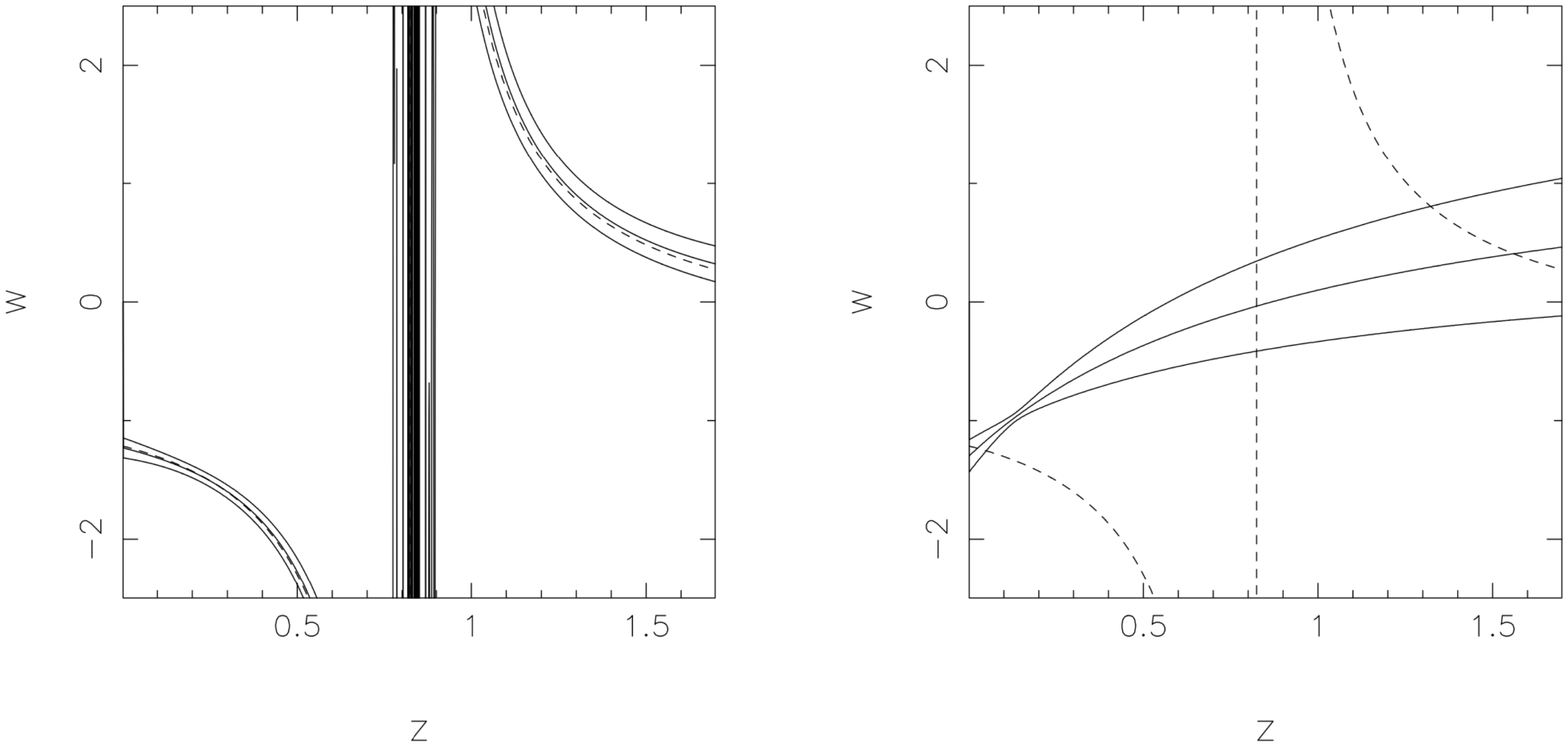}
\vspace{-0.9in}
\caption {Reconstructed equation of state for a braneworld model proposed in~\cite{brane} for 1000 realisations of future-like supernovae data. The left-hand panel shows results using the smoothing method~\cite{smoothing1,smoothing2} while the right panel shows the reconstruction using a dark energy parametrization. Dashed lines represent the actual fiducial model and the solid lines show the reconstructed results using non-parameteric (left) and parameteric (right) approaches. Figure is adopted from~\cite{smoothing1}.}
\label{fig:1}
\end{figure*}

\begin{figure}[!t]
\begin{center}{
\includegraphics[angle=-90,width=\columnwidth]{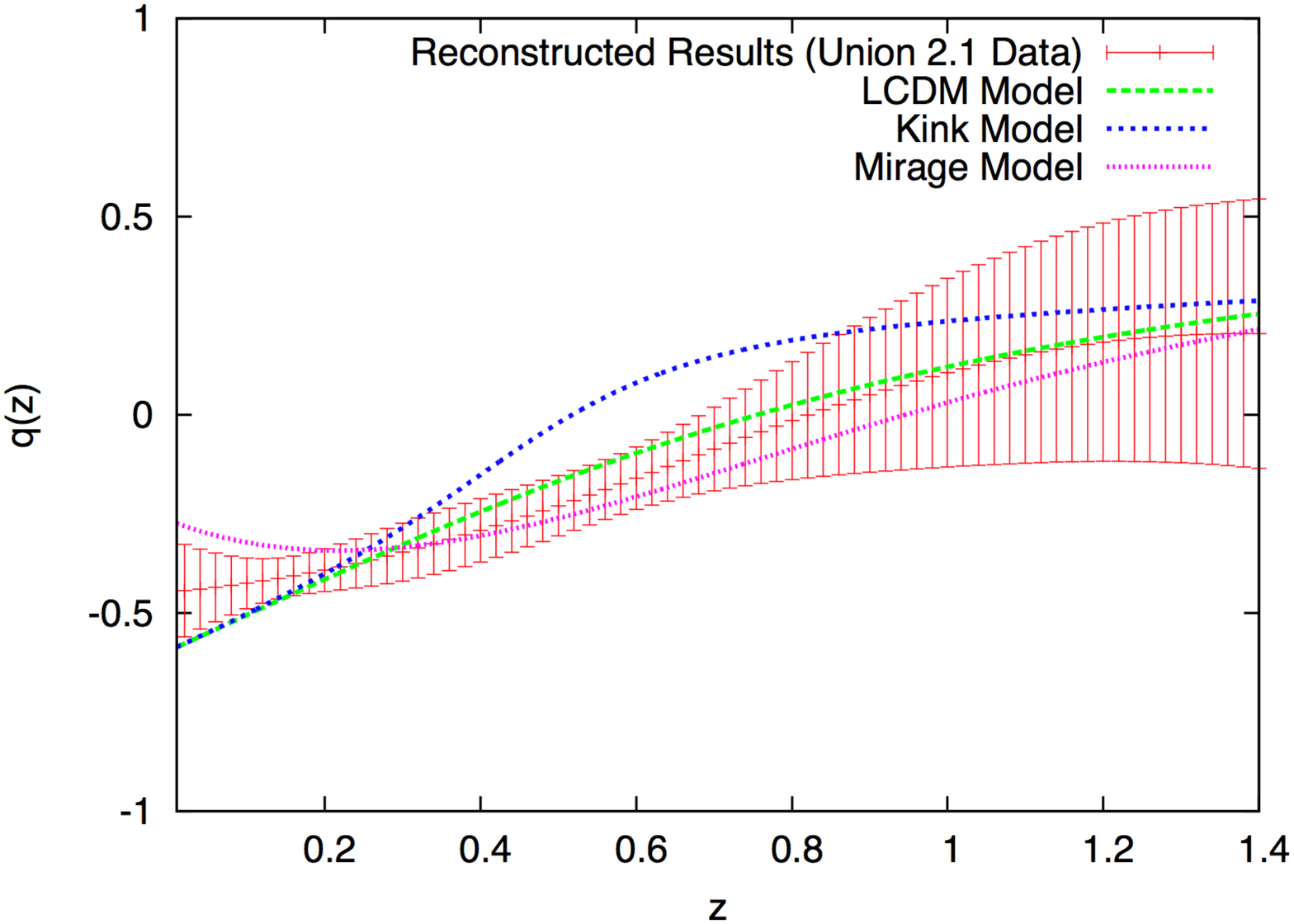}}
\end{center}
\vspace{-.2in}
\caption{GP reconstruction of the deceleration parameter using the Union2.1 data compilation is given by the shaded error band
representing the 68\% confidence level. Three theoretical dark energy model predictions are plotted along side for comparison. Figure is from~\cite{gp2}.}
\label{fig:2}
\end{figure}

In this regard reconstructing the expansion history of the universe and properties of dark energy has become one of the main goals of today’s cosmology to understand our universe and its components. There have been many approaches in last decade proposed to do the reconstruction of the expansion history and one can generalise them in two categories of parametric and non-parametric methods. Parametric methods are viable approaches if we know the actual class-form of the phenomena we are studying and we can use them to put constraints on the parameters of the model. See~\cite{sn_parametric1,sn_parametric2,sn_parametric3,sn_parametric4,sn_parametric5,sn_parametric6} for details of data analysis and methods of parametric reconstruction of the properties of dark energy using supernovae data. However dealing with a phenomena that we have no clear idea about its nature and behaviour, using parametric methods can be misleading since the underlying actual model might not be covered by the assumed parametric form. Dealing with uncertainties in the dispersion of the data adds another complication to the analysis and leave us with no clear way to find this fact that we might have chosen an inappropriate parametric form. This raises the importance of the non-parametric and model independent approaches to find out the behaviour of a phenomena in a more direct way by avoiding parametrizing cosmological quantities~\cite{smoothing1,smoothing2,sn_nonparametric1,sn_nonparametric2,sn_nonparametric3,sn_nonparametric4,sn_nonparametric5,sn_nonparametric6,sn_nonparametric7,sn_nonparametric8,sn_nonparametric9,sn_nonparametric10,gp1,gp2,gp3}. In figure.\ref{fig:1} we can see such a case that using a parametric form of dark energy $w(z) = w_0 + \frac{w_1 z}{1+z}$ ~\cite{cpl1,cpl2} to fit a simulated data based on a brane cosmology model which has a singularity in its effective equation of state~\cite{brane} results to something very much different from the actual fiducial model while a direct non-parametric smoothing method can find the strange feature hidden in the data~\cite{smoothing1,smoothing2}. However one should note that non-parametric approaches have their own shortcomings. For instance, estimation of the errors can be a tricky task in many cases since in some methods one cannot easily assign the degree of freedom in the likelihood analysis. For a review over this subject look at~\cite{de_review}. Recently there have been attempts to combine parametric and non-parameteric methods to reconstruct the expansion history of the universe in order to recover unexpected features of the data as well as defining proper confidence limits. This probably would be an important step towards model independent dark energy reconstruction and resolving dark energy parametrization problem~\cite{crossing_2,crossing_3,gp1,gp2}. In figure.\ref{fig:2} we can see the reconstructed results for the deceleration parameter $q(z)=(1+z)\frac{H'(z)}{H(z)}-1$ ($H(z)$ is the Hubble parameter and $H'(z)$ is its derivative with respect to redshift) using Gaussian Processes (GP) method~\cite{gp2} applied on Union 2.1 supernovae data~\cite{supernovae3} along with theoretical predictions of some dark energy models.  

\section{Cosmographic Degeneracies}
\begin{figure}[!t]
\begin{center}{
\includegraphics[angle=-90,width=\columnwidth]{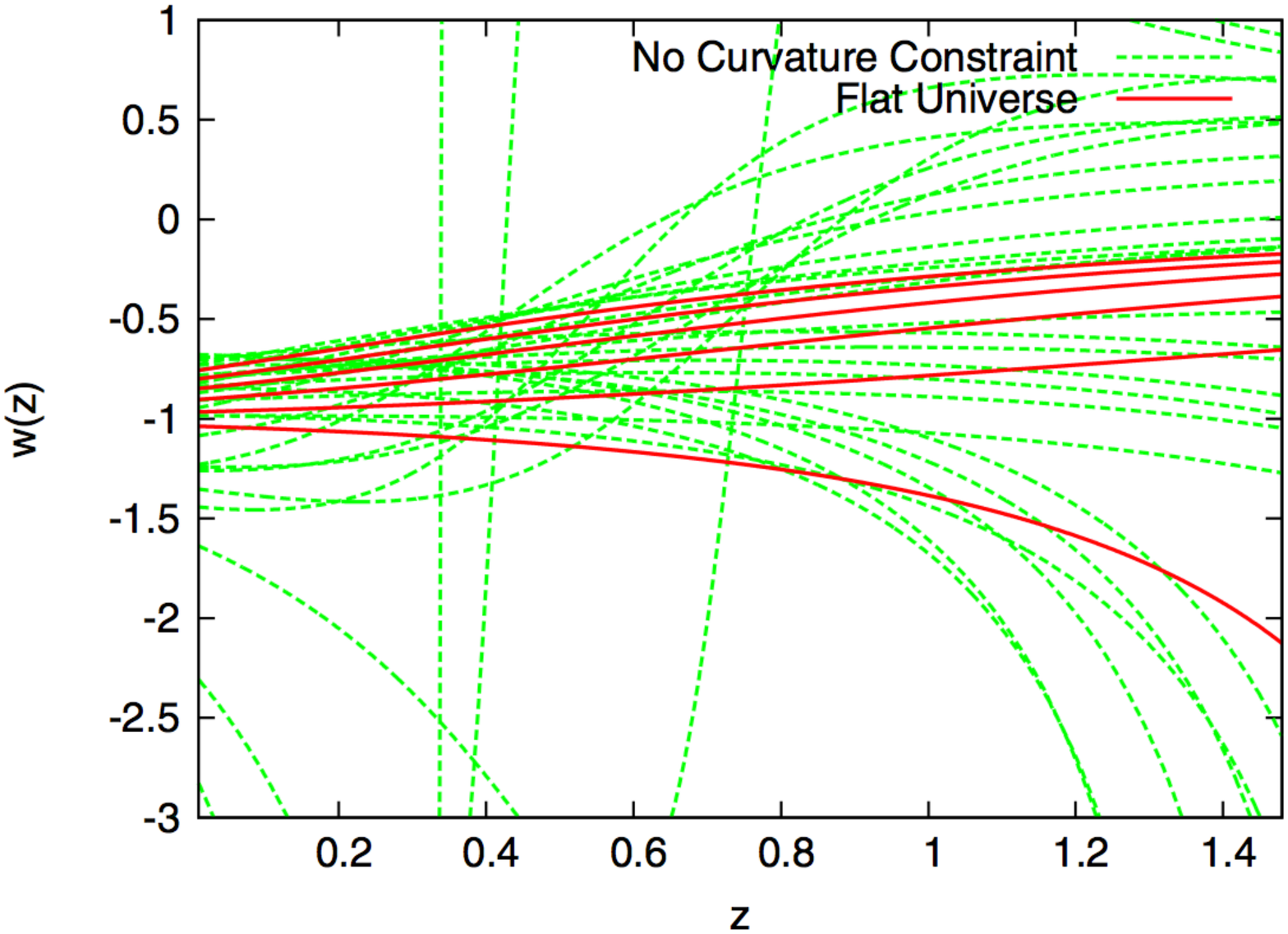}}
\end{center}
\vspace{-.2in}
\caption{ Non-exhaustive sample of $w(z)$ for different points in the $\Omega_m$-$\Omega_{de}$ parameter space that match the $\Lambda$CDM distances exactly out to $z_{max}=1.5$ and satisfy physical conditions at late universe. Light (green) lines represent the results with curvature allowed to be non-flat. Dark (red) lines restrict to the zero-curvature case. The assumed true model is a spatially flat $\Lambda$CDM model with $\Omega_m=0.28$. Figure is adopted from~\cite{CD}}
\label{fig:3}
\end{figure}

 Cosmological observations, especially over the past decade, have made great strides in constraining the energy-density fractions for each of the energy components of the universe. However, as first pointed out in~\cite{maor}, an incorrect prior for the equation of state of dark energy can lead to gross misrepresentations of reality. The same applies to the value of the matter density. In fact in addition to matter (baryonic and dark matter) and small contributions by radiation and neutrinos, there is an (effective) dark energy associated with the accelerated expansion and possibly an (effective) curvature energy associated with deviation from spatial flatness. Because all the energy densities enter into the Hubble expansion rate, which then determines the distance-redshift relation, degeneracies exist between the components such that more of one can compensate for less of another. Since they evolve differently with redshift, however, each characterised by its own equation of state parameter ($0$ for matter, $-1/3$ for curvature, $w(z)$ for dark energy), one expects that observations over a sufficiently wide redshift range give leverage to break the degeneracies. This expectation has been explored for restricted scenarios of matter and dark energy densities (e.g. ~\cite{Huey,smoothing1,smoothing2,smoothing3}) and curvature and dark energy densities (e.g.~\cite{dek1,dek2}), and non-parametrically from the observations through redshifts bins of dark energy. There have been also works to look at how the diversity of models translates to dispersion in observables~\cite{mort1,mort2}. One can also exhaustively investigate the freedom around the concordance model caused by degeneracies when we allow for matter, curvature, and dark energy with no a priori restriction on its equation of state. In fact if we restrict ourselves to late-time observations since we have no knowledge of dark energy behaviour at early times, (e.g. is there early dark energy affecting the cosmic microwave background (CMB)?), we can use purely geometric distance measurements, examining how the degeneracies are broken as the data quality and redshift range improve. It is in fact surprising to understand that when we allow  matter, spatial curvature, and unrestricted dark energy to contribute to the distance-redshift relation, even when perfectly matching the distances out to $z_{max}=1.5$ for a flat $\Lambda$CDM model with a given matter density, a substantial region of the density parameter space remains degenerate with the true model~\cite{CD,luca}. This implies that we cannot assume that the cosmological constant describes the dark energy through such distance measurements alone. Imposing other low-redshift constraints, such as basic consistency conditions on the radius of curvature of closed universes and positivity of the dark energy density, and observational criteria such as a minimum age of the universe and a simple lower bound on the total growth factor for large-scale structure, still leaves considerable freedom for the curvature and dark energy contributions. It is interesting to see how our ignorance of the nature of dark energy and the geometric curvature of space diffuses the strength of evidence for the cosmological constant model from distance measurements. The true universe may be much more complicated, and yet perfectly consistent with cosmography, than this highly restricted model. In figure.\ref{fig:3} we can see how different dark energy models with non-restricted form of equation of state and assuming different curvatures can be degenerate to each other and $\Lambda$CDM model up to an indistinguishable level. Considering the growth data as a complementary information to the cosmological distance data would be certainly very much important to distinguish between various models, in particular to differ between physical and geometrical dark energy models, but it would not be enough to break all the theoretical degeneracies~\cite{gp3,growth}.

\section{Falsifying Cosmological Constant}
Considering the cosmographic degeneracy and difficulties of reconstructing dark energy, taking in to account the quality of cosmological data, we may reach to a conclusion that trying to estimate the equation of state of dark energy to understand this mysterious component of our universe might not be so much plausible. All analysis available in the literature putting tight constraints on the equation of state of dark energy are in fact based on many assumptions and within some particular parametric frameworks. Here, we may change our strategy to concentrate on a more realistic problem. We can try to put the power of the data and find a suitable statistical framework to test and falsify cosmological constant as one of the main aspects of the standard model of cosmology. We should realise that finding any deviation from cosmological constant would result to ruling out the standard model of cosmology and this by itself can be a very important achievement and a breakthrough in the field. So litmus tests of $\Lambda$ have special importance in cosmology and they should follow some particular characteristics. First of all, these diagnostics of cosmological constant should be insensitive to cosmographic degeneracy. In other words we should be able to derive them with minimum knowledge of the individual components of the energy density of the universe. Second, we should be able to derive them directly from cosmological observables without any model assumption or parameterization. Following this policy and keeping in mind these important and crucial characteristics, we have been able to introduce few different diagnostics of cosmological constant and in the following we discuss two of the most important ones: $Om$ and $Om3$. $Om$ diagnostic defined by~\cite{om,om_chris}:

\beq
Om(z) \equiv \frac{h^2(z)-1}{(1+z)^3-1}~,
\label{eq:om}
\eeq
where
\beq
h^2 = \frac{H^2(z)}{H_0^2}=\om (1+z)^3 + \Omega_{\rm DE}~,
\eeq
and
\beq
\Omega_{\rm DE}=(1-\om) \exp{\left\lbrace 3 \int_0^{z}\frac{1+w(z')}{1+z'} dz'\right\rbrace }~.
\label{eq:hubble_recon}
\eeq

\begin{figure}[!t]
\begin{center}{
\includegraphics[angle=-90,width=\columnwidth]{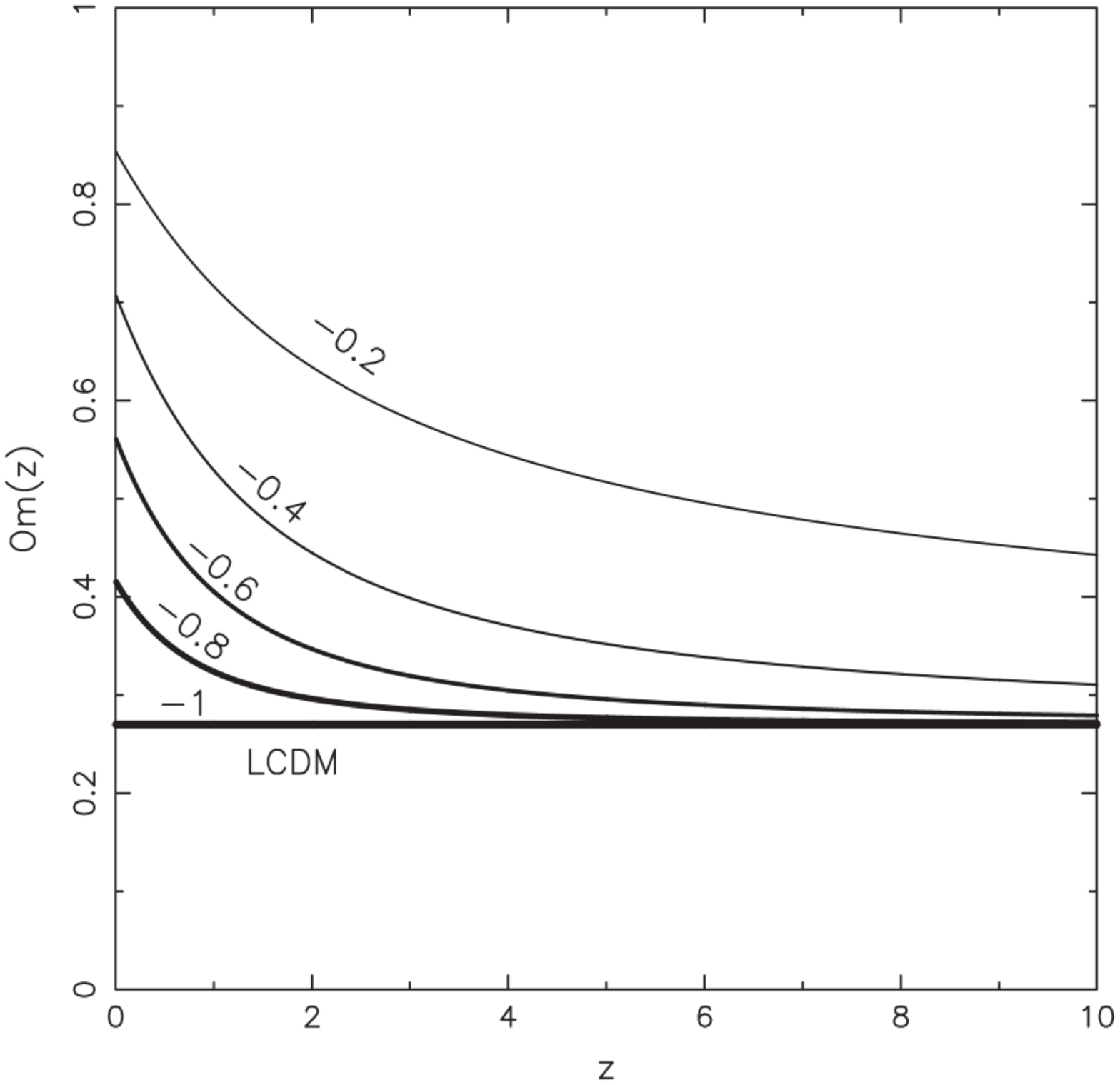}}
\end{center}
\vspace{-.2in}
\caption{ The $Om$ diagnostic is shown as a function of redshift
for DE models with $\om=0.27$ and $w= -1, -0.8, -0.6, -0.4, -0.2$ (bottom to top).
For Phantom models (not shown) $Om$ would have the opposite curvature. Figure is from~\cite{om}.}
\label{fig:om}
\end{figure}

 $h^2(z)$ represents the expansion history of a spatially flat FLRW universe with scale factor $a(t)$ and Hubble parameter $H(z)\equiv \dot a/a$. Important characteristic of $Om$ diagnostic is the fact that it is constant only for flat $\Lambda$CDM model. In contrast to $w(z)$ and the deceleration parameter $q(z) \equiv
-{\ddot a}/aH^2$, the $Om(z)$ diagnostic depends upon no higher
derivative of the luminosity distance than the first one. Therefore, it is less sensitive to observational errors than
either $w$ or $q$. $Om$ is also distinguished by the fact that
$Om(z) = \Omega_m$ for $\Lambda$CDM. In other words, $Om$ is unevolving {\em only for $\Lambda$CDM} and for all other dark energy models the value of $Om(z)$ is redshift dependent.
$Om$ is very useful in establishing the properties of DE. For an
unevolving EOS: $1+w \simeq [Om(z) - \om](1-\om)^{-1}~$ at $z \ll
1$, consequently a larger $Om(z)$ is indicative of a larger $w$;
while at high $z$, $Om(z) \to \om$, as shown in figure.\ref{fig:om}.

We should emphasise that the $Om$ diagnostic depends only upon the expansion history, $h(z)$, so if we can derive $h(z)$ from any cosmological observation we can test cosmological constant hypothesis. $Om$ has been quite successful to become a key parameter in cosmological reconstruction and even major cosmological surveys have used it in last couple of years to represent their results and show consistency of their data with spatially flat $\Lambda$CDM model~\cite{lss_obs1,lss_obs3}.

Though this seems to be quite straight forward to derive $h(z)$  using supernovae data through smoothing or other non-parametric approaches, deriving $h(z)$ from other cosmological observations is not that easy. Note that to derive $h(z)$ from cosmological observations rather than supernovae data (where $H_0$ acts as a nuisance parameter) we need both $H(z)$ and $H_0$ to construct $h(z) =\frac{H(z)}{H_0}$ and uncertainties in $H(z)$ and $H_0$ will add up in this case.

It is widely appreciated that large galaxy surveys, and the resulting Baryon Acoustic Oscillation (BAO) data, hold enormous promise for probing the expansion history of the universe. In defining $Om3$ our focus was on BAO data, with current as well as future observations in mind. The BAO observable (standard ruler) is:
\beq
d(z) = \frac{r_s(z_{\rm CMB})}{D_V(z)}~,
\eeq
where $r_s(z_{\rm CMB})$ is the
 sound horizon marking the decoupling of CMB photons from baryons,
and
\beq
D_V(z) 
= \left [\frac{D_L(z)^2}{(1+z)^2}\frac{cz}{H(z)}\right ]^{1/3} = \left [\frac{D_L(z)^2}{(1+z)^2}\frac{cz}{h(z)H_0}\right ]^{1/3} ~,
\label{eq:DV}
\eeq
$D_L(z)$ being the  luminosity distance.

Clearly, in order to determine $h(z)$ (hence $Om$) from a knowledge of
$d(z)$ we need to know
$r_s(z_{\rm CMB})$, $H_0$, $d(z)$ and $D_L(z)$. In~\cite{smoothing3} it had been shown that error propagation in these four quantities makes it rather difficult to reconstruct $h(z)$ and $Om(z)$ in a precise model independent manner
 (see for instance fig.4 of~\cite{smoothing3}).

This issue was the main motivation to look for a new diagnostic that depends on fewer observables.

\begin{figure}[!t]
\begin{center}{
\includegraphics[angle=-90,width=\columnwidth]{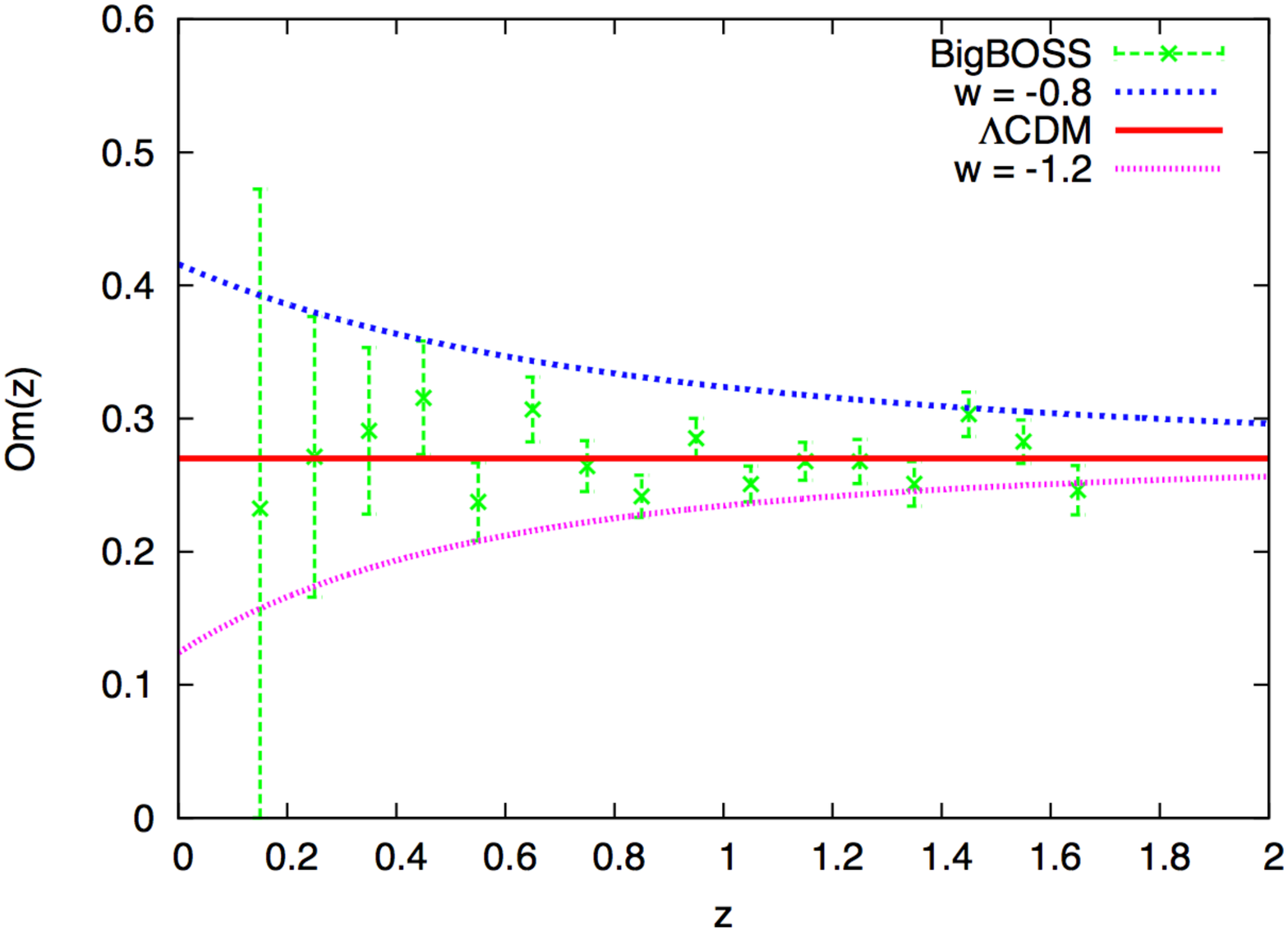}
\includegraphics[angle=-90,width=\columnwidth]{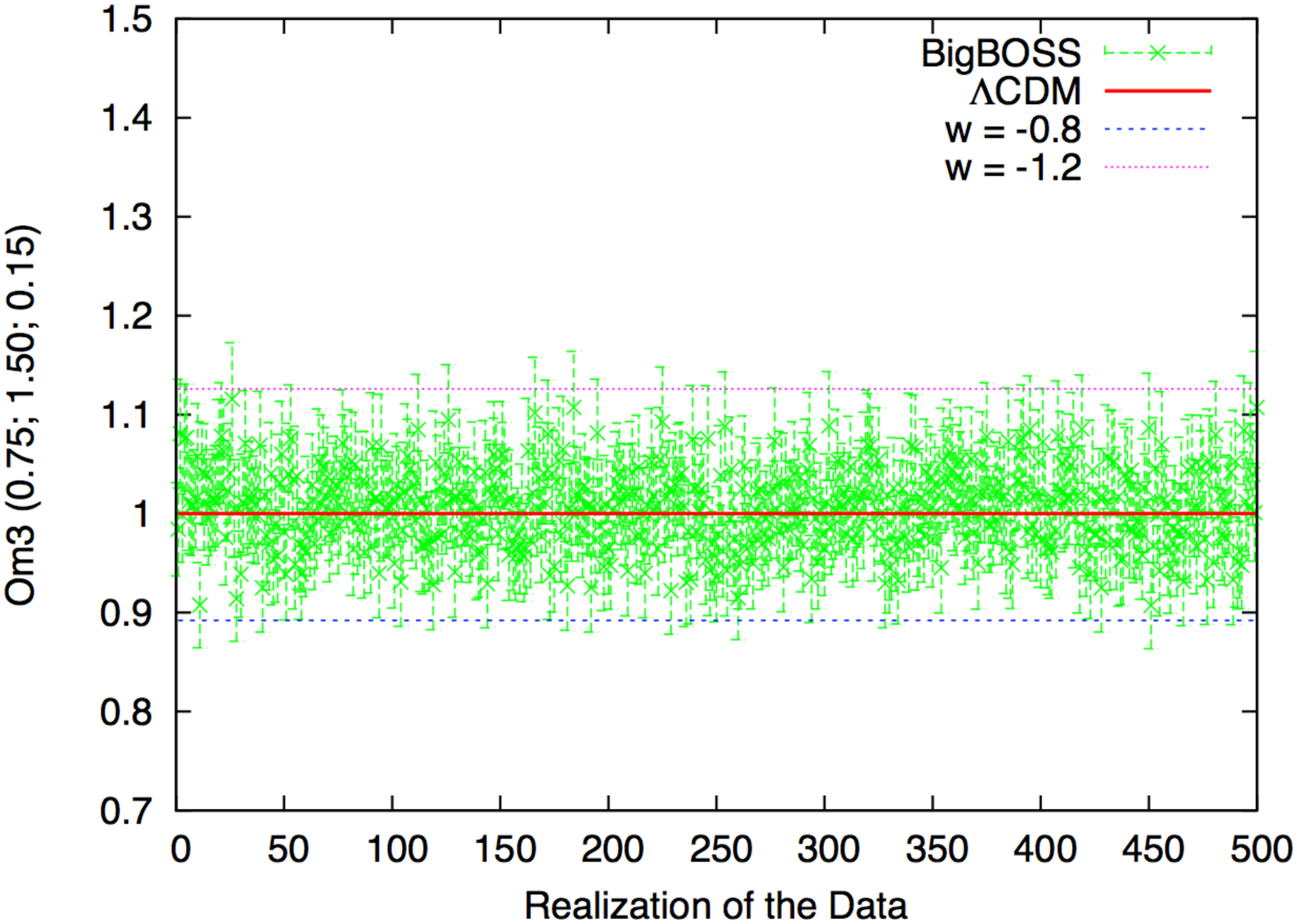}}
\end{center}
\vspace{-.2in}
\caption{ Top panel:The $Om$ diagnostic is reconstructed for a single simulated realisation of the planned BigBOSS experiment~\cite{bigboss} assuming a fiducial $\Lambda$CDM cosmology. A determination of $h(z)$, and hence $Om(z)$, from future BAO experiments can clearly help
distinguish between rival models of dark energy. Note that this determination  is based on values of
$H_0$ with $2\%$ uncertainty expected by the time BigBOSS becomes operational. Bottom panel: $Om3$ derived using simulated realisations of the
BigBOSS experiment~\cite{bigboss} assuming a fiducial $\Lambda$CDM cosmology. Horizontal lines
represent different dark energy models with (top-down) $w=-0.8$,
 $w=-1.0$, $w=-1.2$. $\Omega_{m}=0.27$ is assumed for all models. Note that the determination of $Om3$ requires {\em minimal cosmological assumptions} since one does not require a background model to estimate $\Omega_{m}$, $H_0$, or the distance to the last scattering surface. Figures are adopted from~\cite{om3}.}
\label{fig:om3}
\end{figure}

The resultant diagnostic $Om3$ is defined by~\cite{om3}:
\beq
Om3(z_1;z_2;z_3) = \frac{H(z_2;z_1)^2-1}{x_2^3-x_1^3}\bigg/
\frac{H(z_3;z_1)^2-1}{x_3^3-x_1^3},
\label{eq:om3a}
\eeq 
where $x = 1+z$ and
\beq
H(z_i;z_j)
= \frac{z_i}{z_j}\left [ \frac{D_L(z_i)(1+z_j)}{D_L(z_j)(1+z_i)}\right ]^2\left [\frac{d(z_i)}{d(z_j)}\right ]^3~.
\label{eq:fracH2}
\eeq
Note that $Om3$ depends {\em only on} $D_L$ and $d(z)$ and that
{\bf there is no dependence on  $r_s(z_{\rm CMB})$ and $H_0$}.
Thus the number of independent observables has been reduced from four to just two,
which makes $Om3$ independent of assumptions that go into the
determination of $r_s(z_{\rm CMB})$ and insensitive to uncertainties in $H_0$. $Om3$ is equal to unity only for spatially flat $\Lambda$CDM model and any deviation from one represent inconsistency with $\Lambda$ dark energy. In fingure.\ref{fig:om3} we see the expected results for $Om$ and $Om3$ diagnostics from the future BigBOSS large scale structure experiment~\cite{bigboss}.

\section{Conclusion}
I have briefly reviewed two strategic approaches of studying dark energy, reconstructing dark energy and falsifying dark energy models. Difficulties of reconstructing the expansion history of the universe and the properties of dark energy were discussed and we had also expressed the problem of cosmographic degeneracy which would not allow us to confidently distinguish between cosmological models with high certainty at the present and near future. We have then emphasised on importance of a different approach to confront a  more realistic and affordable problem: falsifying cosmological constant. Cosmological constant or $\Lambda$-term as dark energy is one of the important aspects of the standard cosmological model and finding any deviation from it would result to a break through in theoretical physics and ruling out the standard concordance model of cosmology. We have explained two diagnostics of dark energy which are suitably designed to test the $\Lambda$ term using direct observables. While $Om$ diagnostic can be trivially applied using supernovae data, $Om3$ is tailored specifically to be used on large scale structure data and BAO direct observables. We have shown effectiveness of these two important diagnostics in the near future of cosmology and how they can test the standard model without assuming any parameterization, without being sensitive to the priors of individual quantities and most importantly, by only using the direct observables.




\section{Acknowledgment}
A.S. acknowledges the Max Planck Society (MPG), the Korea Ministry of Education, Science and Technology (MEST), Gyeongsangbuk-Do and Pohang City for the support of the Independent Junior Research Groups at the Asia Pacific Center for Theoretical Physics (APCTP).

\nocite{*}
\bibliographystyle{elsarticle-num}
\bibliography{martin}



\end{document}